%% file: main.tex
\documentclass[sigconf]{acmart}

\usepackage[utf8]{inputenc}
\usepackage{booktabs,multirow}
\usepackage{enumitem}
\usepackage{color}
\usepackage{rotating}
\usepackage{amsmath}
\usepackage{array}
\usepackage{enumitem}
\usepackage[linesnumbered,algoruled,boxed,lined]{algorithm2e}
\usepackage{balance}
\usepackage[short]{optidef}
\usepackage{multirow}
\usepackage{graphicx} % for rotatebox
\usepackage{pifont}% http://ctan.org/pkg/pifont
\usepackage{adjustbox}
\usepackage{bm}
\usepackage{algpseudocode}
\newcolumntype{P}[1]{>{\centering\arraybackslash}p{#1}}

\newcommand{\ie}{{\it i.e.}}
\newcommand{\eg}{{\it e.g.}}
\newcommand{\ul}{\underline}{}

\copyrightyear{2024}
\acmYear{2024}
\setcopyright{rightsretained}
\acmConference[CIKM '24]{Proceedings of the 33rd ACM International Conference on Information and Knowledge Management}{October 21--25, 2024}{Boise, ID, USA}
\acmBooktitle{Proceedings of the 33rd ACM International Conference on Information and Knowledge Management (CIKM '24), October 21--25, 2024, Boise, ID, USA}
\acmDOI{10.1145/3627673.3679960}
\acmISBN{979-8-4007-0436-9/24/10}

\makeatletter
\gdef\@copyrightpermission{
  \begin{minipage}{0.3\columnwidth}
   \href{https://creativecommons.org/licenses/by-nc-sa/4.0/}{\includegraphics[width=0.90\textwidth]{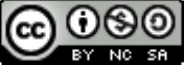}}
  \end{minipage}\hfill
  \begin{minipage}{0.7\columnwidth}
   \href{https://creativecommons.org/licenses/by-nc-sa/4.0/}{This work is licensed under a Creative Commons Attribution-NonCommercial-ShareAlike International 4.0 License.}
  \end{minipage}
  \vspace{5pt}
}
\makeatother

\begin{document}

\title{MARS: Matching Attribute-aware Representations for Text-based Sequential Recommendation}
    
\author{Hyunsoo Kim} \authornote{Both authors contributed equally to this research.}
\affiliation{
  \institution{Sungkyunkwan University}
  \city{Suwon}
  \country{Republic of Korea}}
\email{khs1778@skku.edu}

\author{Junyoung Kim} \authornotemark[1]
\affiliation{
  \institution{Sungkyunkwan University}
  \city{Suwon}
  \country{Republic of Korea}}
\email{junyoung44@skku.edu}

\author{Minjin Choi}
\affiliation{
  \institution{Sungkyunkwan University}
  \city{Suwon}
  \country{Republic of Korea}}
\email{zxcvxd@skku.edu}

\author{Sunkyung Lee}
\affiliation{
  \institution{Sungkyunkwan University}
  \city{Suwon}
  \country{Republic of Korea}}
\email{sk1027@skku.edu}

\author{Jongwuk Lee}\authornote{Corresponding author}
\affiliation{
  \institution{Sungkyunkwan University}
  \city{Suwon}
  \country{Republic of Korea}}
\email{jongwuklee@skku.edu}




\input{body/0_abstract}

\maketitle

\input{body/1_introduction}
\input{body/2_method}
\input{body/3_results}
\input{body/4_conclusion}

\begin{acks}
    This work was supported by Institute of Information \& communications Technology Planning \& Evaluation (IITP) grant and National Research Foundation of Korea (NRF) grant funded by the Korea government (MSIT) (No. RS-2019-II190421, IITP-2024-2020-0-01821, RS-2022-II220680, and RS-2021-II212068).
\end{acks}

\bibliographystyle{ACM-Reference-Format}
\balance
\bibliography{citations}

\end{document}

%% file: body/0_abstract.tex
\begin{abstract}
Sequential recommendation aims to predict the next item a user is likely to prefer based on their sequential interaction history.
Recently, text-based sequential recommendation has emerged as a promising paradigm that uses pre-trained language models to exploit textual item features to enhance performance and facilitate knowledge transfer to unseen datasets.
However, existing text-based recommender models still struggle with two key challenges: (i) representing users and items with multiple attributes, and (ii) matching items with complex user interests.
To address these challenges, we propose a novel model, \emph{\textbf{M}atching \textbf{A}ttribute-aware \textbf{R}epresentations for Text-based \textbf{S}equential Recommendation (MARS)}.
MARS extracts detailed user and item representations through \textit{attribute-aware text encoding}, capturing diverse user intents with multiple attribute-aware representations. It then computes user-item scores via \textit{attribute-wise interaction matching}, effectively capturing attribute-level user preferences.
Our extensive experiments demonstrate that MARS significantly outperforms existing sequential models, achieving improvements of up to 24.43\% and 29.26\% in Recall@10 and NDCG@10 across five benchmark datasets.
\footnote{Code is available at \href{https://github.com/junieberry/MARS}{https://github.com/junieberry/MARS}.}
\end{abstract}

%% file: body/1_introduction.tex
\section{Introduction}\label{sec:introduction}

\input{Figures/fig_motivation}

Sequential recommendation~\cite{FangGZS19Survey, HidasiKBT15GRU4Rec, KangM18SASRec, FangZSG20SeqSurvey, WangHWCSO19SeqSurvey, ChenXZT0QZ18MANN} aims to deliver consecutive items to users based on a sequential form of their past interaction history.
Due to the dynamic evolution of user behavior over time, they are vital in various commercial web applications, \eg, Amazon~\cite{LindenSY03Amazon}, YouTube~\cite{CovingtonAS16}, and Spotify~\cite{ChenLSZ18Spotify}. 

Conventional ID-based models~\cite{KangM18SASRec, HidasiKBT15GRU4Rec, LiZZYCSKN22RecGURU} represent items as explicit IDs. They randomly initialize vectors for item embeddings and optimize them with sequential user-item interactions through neural encoders, such as CNNs~\cite{TangW18caser}, MLPs~\cite{ZhouYZW22FMLPRec}, RNNs~\cite{HidasiKBT15GRU4Rec, LiRCRLM17NARM}, GNNs~\cite{ChangGZHNSJ021SURGE, WuT0WXT19SRGNN}, and Transformers~\cite{KangM18SASRec, WuLHS20SSEPT, SunLWPLOJ19BERT4Rec, DuSZWSLL022CBIT, QiuHYW22DuoRec, XieSLWGZDC22CL4SRec}.
However, they struggle to learn item representations with sparse interactions, \ie, cold-start items.
Some studies~\cite{ZhouWZZWZWW20S3Rec, ZhangZLSXWLZ19FDSA, XieZK22DIFSR, Liu21NOVA, YuanDTSZ21ICAISR} have exploited item metadata as side information to compensate for the lack of collaborative signal.
However, these approaches suffer from the fundamental limitation of handling new items or new attributes.

Recent studies~\cite{HouMZLDW22UniSRec, TangWZZZL23MIRACLE, LiWLFSSM23Recformer, LiuMXLYL0023TASTE, Hao21ZesRec} have proposed text-based sequential recommendation that uses pre-trained language models (PLMs).
These methods perform matching by encoding textual representations of the user/item into embedding vectors.
They offer two advantages: (i) handling unseen items and transferring knowledge to new datasets without additional training, and (ii) enriching item representations with real-world knowledge.
Nevertheless, existing models struggle to capture nuanced user interests and item features, as they represent user sequences and items as single vectors.
MIRACLE~\cite{TangWZZZL23MIRACLE} attempts to enhance user sequence representations with multiple vectors, but is limited by a fixed number of interests and single-vector item representations.

Figure ~\ref{fig:motivation} illustrates a scenario where two users with similar purchase histories receive distinct item recommendations based on their individual attribute-level preferences.
For instance, User A may be recommended a ``Magic Keyboard'' from ``Apple'' given their purchase history of ``Magic Mouse'' and ``Keyboard.''
In contrast, User B's purchase of ``Gaming Headset'' and ``Logitech'' products suggests an interest in gaming, which may lead to a recommendation for a ``Gaming Mouse'' following their recent ``Mouse'' purchase.
These scenarios emphasize the need to identify user interests by combining purchase history in an attribute-wise manner, as a single vector representation of the sequence cannot capture preferences for specific attributes like ``Logitech'' or ``Keyboard.''

In light of these observations, we establish two challenges: 
(i) \emph{How do we represent a user/item to represent multiple attributes?}
(ii) \emph{How do we calculate the matching score between the user and the item considering complex user interests?}
To address these challenges, we propose a novel text-based sequential recommender model, namely \emph{\textbf{M}atching \textbf{A}ttribute-aware \textbf{R}epresentations for \textbf{S}equential Recommendation (MARS)}, which learns fine-grained user preferences by leveraging textual information of item attributes as shown in Figure~\ref{fig:model_architecture}.
Specifically, we obtain \textit{multiple attribute-aware representations} by encoding the textual attributes of user sequence and item with PLMs, preserving fine-grained semantics.
The user-item score is then computed via \textit{attribute-wise item matching} using multiple representations. The score is calculated for each attribute between user sequences/items, which can finely represent the attribute-level preference.
This enables MARS to capture the intricate user-item relationship between the fine-grained representations.
Extensive experimental results on five benchmark datasets show that MARS outperforms existing sequential recommendation models by up to 24.43\% and 29.26\% in Recall@10 and NDCG@10, respectively.

%% file: Figures/fig_motivation.tex
\begin{figure}[t]
\includegraphics[width=1\linewidth]{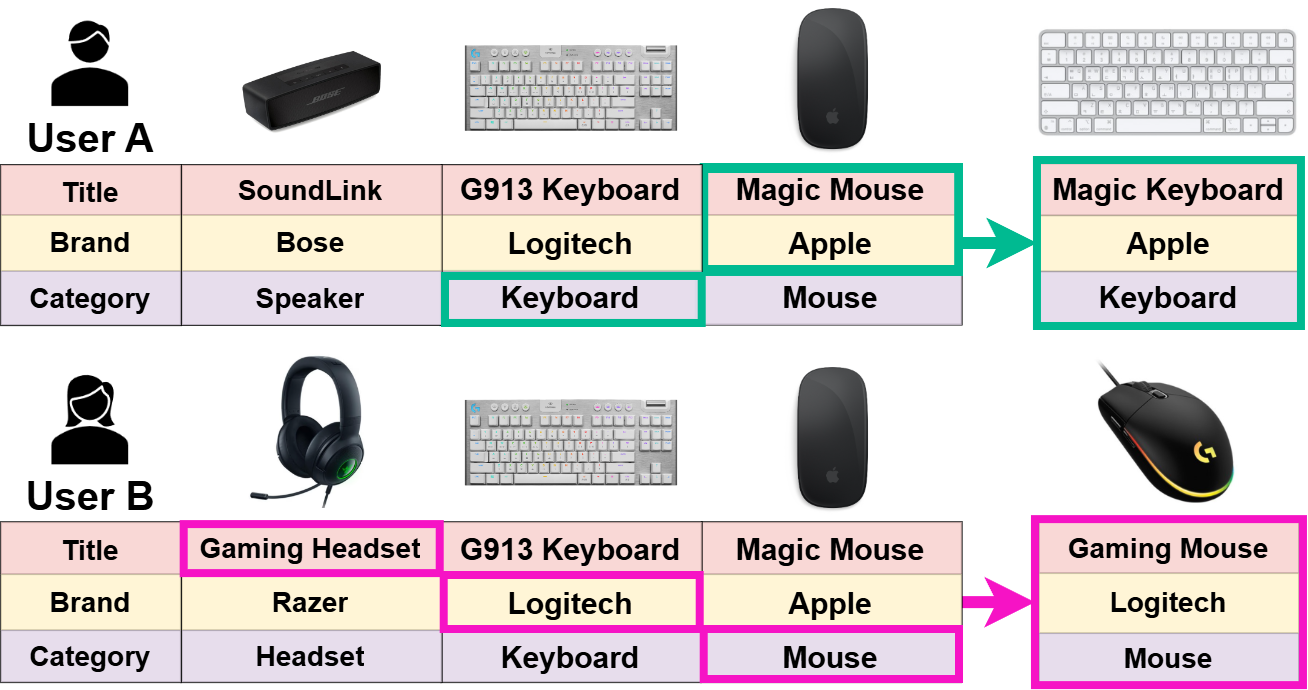}
% \vspace{-5mm}
\caption{A motivating example of two user sequences. The green and pink boxes represent different user interests that are highly relevant to target items.}\label{fig:motivation}
\vspace{-6mm}
\end{figure} 

%% file: body/2_method.tex
\section{Proposed Model}\label{sec:model}

\input{Figures/fig_model_architecture}

\subsection{Text-based Sequential Recommendation}\label{sec:text_background}

\noindent
\textbf{Problem definition.}
Let $\mathcal{U}$ and $\mathcal{I}$ denote a set of users and a set of items. 
The interaction history of user $u \in \mathcal{U}$ is denoted by $s=(i_1, i_2, \dots, i_n)$, where $i_t \in \mathcal{I}$ refers to the item interacted at the $t$-th position, and $n$ indicates the number of items in the user sequence $s$.
Text-based sequential recommender models utilize \emph{textual item information} to predict the next item $i_{n+1}$ that a user will likely engage with, based on their historical sequence $s$.

\noindent
\textbf{Overall process.}
Following \cite{HouMZLDW22UniSRec,TangWZZZL23MIRACLE,LiWLFSSM23Recformer}, each item $i \in \mathcal{I}$ is represented by a list of textual attributes $T_i = (a_{i,1}, \dots, a_{i,m})$, where $m$ is the number of item attributes. 
The attribute $a_{i,j}$ is represented in a key-value format: $a_{i,j} = (k_{i,j}, v_{i,j})$, where $k_{i,j}$ and $v_{i,j}$ indicate the attribute name (\eg, Category) and value \ (\eg, Keyboard), respectively.
$k_{i,j}$ and $v_{i,j}$ are joined to represent $a_{i,j}$ in text (\eg, `Category: Keyboard').
User sequence can be represented in text as $T_s = (T_n, \dots, T_1) = ([a_{n, 1}, \dots, a_{n, m},  a_{n-1, 1}, \dots, a_{1, m}])$, where $a_{t, j}$ is the textual description of the $j$-th attribute of the item $i_t$.
Note that we input the sequence in reverse order to prevent recent items from being truncated~\cite{LiWLFSSM23Recformer, TangWZZZL23MIRACLE}.

Finally, given a user sequence $T_s$ and an item $T_i$, the recommendation score $S(s, i)$ is calculated as follows. 
\begin{equation} \label{eq:recommendation_score}
    S(s, i) = \text{sim}(\text{Enc}(T_s), \text{Enc}(T_i)),
\end{equation}
where $\text{Enc}(\cdot)$ represents the text encoder, and $\text{sim}(\cdot, \cdot)$ is a scoring function. In the following sections, we introduce our proposed model focusing on two key components of text-based sequential recommendation: $\text{Enc}(\cdot)$ in Section~\ref{sec:attribute_encoder} and $\text{sim}(\cdot, \cdot)$ in Section~\ref{sec:attribute_matching}.

% \vspace{-2mm}
\subsection{Attribute-aware Text Encoding}\label{sec:attribute_encoder}
While existing text-based sequential recommenders~\cite{HouMZLDW22UniSRec, LiWLFSSM23Recformer} represent the user sequence/item as a single coarse-grained vector using PLM,
we obtain \emph{attribute-aware multiple representations} to explicitly reflect the diverse preferences and aspects of user sequences and items.

\noindent
\textbf{Item encoder.}
The attribute $j$ of item $i$ is tokenized as follows:
\begin{equation}
    X_{i,j} = \text{Tokenize}\left(a_{i,j}\right) = \left(x_{i,j}^{(1)}, \dots, x_{i,j}^{(L')}\right),
\end{equation}
where $L'$ represents the length of tokenized $a_{i,j}$.
The text representation $X_i$ of item $i$, can be obtained by concatenating the tokenized attributes of the item:
\begin{equation}
    X_i = \left(X_{i,1}; \dots; X_{i,m}\right) = \left(x_1, x_2, \dots, x_L\right),
\end{equation}
where $L$ is the length of $X_i$.
By prefixing $X_i$ with a \texttt{[BOS]} token and passing it through the PLM, we obtain the contextualized vector $\mathbf{o} \in \mathbb{R}^{d}$ for each token.
\begin{equation}
    \mathbf{O}_i = \left[\mathbf{o}_\texttt{[BOS]}, \mathbf{o}_1, \dots, \mathbf{o}_{L}\right] = \text{PLM}\left(\left[\texttt{[BOS]}; X_i\right]\right),
\end{equation}
where $d$ is the hidden dimension size of the PLM, $\mathbf{o}_l$ is the last hidden state of the input token $x_l$, and $\mathbf{O}_i$ is the list of hidden states for item $i$.
We use Longformer~\cite{05150Longformer} to handle long sequence input.

Contrary to existing studies~\cite{HouMZLDW22UniSRec, LiWLFSSM23Recformer} that represent user with a single coarse-grained vector, \eg, $\mathbf{o}_\texttt{[BOS]}$,
we utilize fine-grained representations through attribute-aware multiple vectors.
First, to alleviate matching complexity, we employ a linear layer to reduce the dimension of $\mathbf{o}_l$ from $d$ to $d'$ by setting $d'<d$:
\begin{equation}
    \mathbf{O}'_i = \left[\mathbf{o}'_{\texttt{[BOS]}}, \mathbf{o}'_1, \dots, \mathbf{o}'_L\right] = \left[ \left\{ \mathbf{W}^\top \mathbf{o}+\mathbf{b} \mid \mathbf{o} \in \mathbf{O}_i \right\} \right]
\end{equation}
where $\mathbf{o}'_l \in \mathbb{R}^{d'}$, weight $\mathbf{W} \in \mathbb{R}^{d \times d'}$, and bias $\mathbf{b} \in \mathbb{R}^{d'}$.

The attribute-wise representation of $a_{i,j}$ is obtained by aggregating the hidden states of the tokens that describe $a_{i,j}$.
Specifically, for each attribute $a_{i,j}$, we compute the representation $\mathbf{h}_{i,j}$ by averaging the hidden states $\mathbf{o}'_l \in \mathbf{O}'_i$ that corresponds to the tokens $x_l \in X_{i,j}$.

\begin{equation} \label{attribute_avgpool_score}
    \mathbf{h}_{i,j} = \text{AvgPool}\left(\{ \mathbf{o}'_l \mid x_l \in X_{i,j}, \ \mathbf{o}'_l \in \mathbf{O}'_i \}\right)
\end{equation}
Finally, $T_i$ is encoded into multiple attribute-aware representations by applying Eq. ~\eqref{attribute_avgpool_score} to $m$ attributes.
\begin{equation} \label{eq:item_encoding}
    \text{Enc}\left(T_i\right) = \left[\mathbf{h}_{i,1}, \dots, \mathbf{h}_{i,m}\right].
\end{equation}

\noindent
\textbf{User encoder.} 
User sequences are encoded into attribute-aware multiple representations to capture various user intents. By tokenizing the text representation of the sequence $T_s$, the input $X_s$ is represented as follows:
\begin{equation}
    \begin{aligned}
        X_s 
        &= \left(X_{n,1}; \dots; X_{n,m}; \dots; X_{1, 1}; \dots; X_{1, m}\right)
    \end{aligned}
\end{equation}
By following the similar process in the item encoder, we then obtain the contextualized vector $\mathbf{h}_{t, j} \in \mathbb{R}^{d'}$ for each item attribute $a_{t, j}$.
\begin{equation}
    \text{Enc}\left(T_s\right) = \left[\mathbf{h}_{n, 1}, \dots, \mathbf{h}_{n, m}, \dots, \mathbf{h}_{1, 1}, \dots, \mathbf{h}_{1, m}\right].
\end{equation}
Note that the encoder $\text{Enc}(\cdot)$ is shared with the item encoder to effectively learn the semantics of attributes in the user/item.

\subsection{Attribute-wise Interaction Matching} \label{sec:attribute_matching}

Since the user sequence and item are represented as multiple vectors (specifically, $n \times m$ and $m$ vectors), computing a matching score is nontrivial.
We devise \textit{attribute-wise interaction matching} to fully account for the relationship between multiple user sequence/item attribute-aware representations.
The score is calculated for each attribute and then aggregated to obtain the final score.
We employ maximum similarity (MaxSim)~\cite{KhattabZ20ColBERT} to dynamically match each item attribute with different aspects of the user sequence, capturing user interest based on candidate item attributes.
For example, in Figure \ref{fig:motivation}, “G913 Keyboard” and “Magic Mouse” appear identically in both sequences, but the attributes that contribute to the user's preference are completely different.

We can modify Eq. ~\eqref{eq:recommendation_score} for attribute-wise interaction matching.
The matching score $S_j(s, i)$ between user sequence $s$ and item $i$ on attribute $j$ is calculated as follows:
\begin{equation} \label{eq:attribute_wise_matching_score}
    S_j\left(s, i\right) = \max_{t \in \{n, \dots, 1\}} \text{cos}\left(\mathbf{h}_{t, j}, \mathbf{h}_{j}\right).
\end{equation}
The final matching score $S(s, i)$ is determined by summing the scores for all attributes:
\begin{equation} \label{eq:total_matching_score}
    S\left(s,i\right) = \sum_{j=1}^m {S_j\left(s, i\right)}.
\end{equation}
In this way, items that achieve the highest matching scores across all attributes are recommended to users.

\subsection{Training and Inference}

Finally, we describe the training and inference process. We utilize the cross-entropy loss following ~\cite{HouMZLDW22UniSRec,TangWZZZL23MIRACLE,LiWLFSSM23Recformer}:
\begin{equation} \label{eq:training_loss}
    \mathcal{L} = - \log\frac{\text{exp}\left(S \left( s,i^{+}\right) / \tau \right)}{\sum_{i \in \mathcal{I}} {\text{exp}\left( S\left( s,i \right) / \tau\right)}},
\end{equation}
where $i^{+}$ is a target item for the given sequence and $\tau$ is a hyperparameter to control the temperature~\cite{HintonVD15distillation} for better convergence~\cite{GuptaGMVS19NISER}.

To avoid prolonging training time, we employ two-stage training as in \cite{LiWLFSSM23Recformer}.
In the initial phase, item representations are updated every epoch until convergence.
Then, we fix the optimal item embedding and resume training to fully optimize the model.
At inference, we precompute attribute-aware representations for all items and calculate recommendation scores for each sequence using Eq.~\eqref{eq:total_matching_score}, serving the top-$K$ items with the highest scores.

\input{Tables/tab-main_table}

%% file: Figures/fig_model_architecture.tex
\begin{figure}[t]
% \vspace{-3mm}
% 1 column version
\centering
\includegraphics[width=1\linewidth]{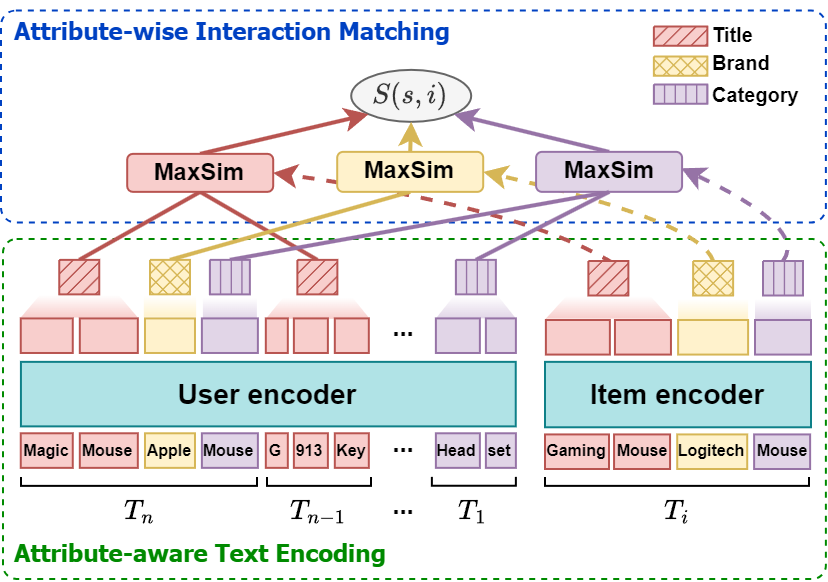} 
\caption{Model architecture of MARS. Each item consists of three attributes, \eg, title, brand, and category. For simplicity, we only display attribute values, omitting their keys.}\label{fig:model_architecture}
\vskip -0.2in
% \vspace{-5mm}
\end{figure}

%% file: Tables/tab-main_table.tex
\begin{table*}[t]\small
\centering
\renewcommand{\arraystretch}{0.8}
\caption{Overall performance. The best results are shown in \textbf{bold}, and the second-best results are \ul{underlined}. "Gain" represents the percentage improvement of MARS over the best competing baseline. `*' indicates statistically significant improvements $(p < 0.01)$ of MARS compared to the best baseline, as determined by a paired $t$-test conducted across 5 experiments.}
\vspace{-2.5mm}
\label{tab:main_table}

\begin{tabular}{c|c|ccc|cc|cccc|c}
\toprule
\multirow{2}{*}{{Dataset}} & \multirow{2}{*}{{Metric}} & \multicolumn{3}{c|}{{ID-based}} & \multicolumn{2}{c|}{{ID + Text}} & \multicolumn{4}{c|}{{Text-based}} & \multirow{2}{*}{{Gain}} \\
 &  & {GRU4Rec} & {SASRec} & {BERT4Rec} & {FDSA} & {S$^3$Rec} & {UniSRec} & {MIRACLE} & {Recformer} & {MARS} &  \\ \midrule
\multirow{2}{*}{{Scientific}} & {R@10} & 0.0562 & 0.1350 & 0.0616 & 0.1160 & 0.0745 & 0.1040 & 0.1006 & {\ul {0.1435}} & \textbf{0.1533*} & 6.83\% \\
 & {N@10} & 0.0353 & 0.0840 & 0.0365 & 0.0827 & 0.0462 & 0.0720 & 0.0829 & {\ul {0.1002}} & \textbf{0.1086*} & 8.38\% \\ \midrule
\multirow{2}{*}{{Pantry}} & {R@10} & 0.0368 & 0.0781 & 0.0441 & 0.0664 & 0.0540 & 0.0548 & 0.0512 & {\ul {0.0840}} & \textbf{0.0928*} & 10.47\% \\
 & {N@10} & 0.0190 & 0.0487 & 0.0289 & 0.0441 & 0.0293 & 0.0386 & 0.0379 & {\ul {0.0535}} & \textbf{0.0591*} & 10.53\% \\ \midrule
\multirow{2}{*}{{Inst.}} & {R@10} & 0.0551 & 0.0934 & 0.0699 & 0.0766 & 0.0723 & 0.0828 & 0.0871 & {\ul {0.0941}} & \textbf{0.1082*} & 14.98\% \\
 & {N@10} & 0.0352 & 0.0632 & 0.0459 & 0.0537 & 0.0447 & 0.0642 & {\ul {0.0720}} & 0.0691 & \textbf{0.0846*} & 17.46\% \\ \midrule
\multirow{2}{*}{{Arts}} & {R@10} & 0.0678 & 0.1342 & 0.1048 & 0.1275 & 0.0823 & 0.1095 & 0.1140 & {\ul {0.1501}} & \textbf{0.1684*} & 12.23\% \\
 & {N@10} & 0.0421 & 0.0908 & 0.0646 & 0.0918 & 0.0581 & 0.0830 & 0.1005 & {\ul {0.1094}} & \textbf{0.1306*} & 19.45\% \\ \midrule
\multirow{2}{*}{{Office}} & {R@10} & 0.0562 & 0.1187 & 0.0985 & 0.0999 & 0.0723 & 0.0990 & 0.1098 & {\ul {0.1263}} & \textbf{0.1571*} & 24.43\% \\
 & {N@10} & 0.0353 & 0.0841 & 0.0701 & 0.0715 & 0.0450 & 0.0810 & {\ul {0.0982}} & 0.0920 & \textbf{0.1270*} & 29.26\% \\
\bottomrule
\end{tabular}
\vspace{-3mm}
\end{table*}

%% file: body/3_results.tex
\section{Experiments}\label{sec:exp_overall}
\subsection{Experimental Setup}\label{sec:setup}

\noindent
\textbf{Datasets.} 
We used the Amazon review dataset\footnote{\url{https://cseweb.ucsd.edu/~jmcauley/datasets/amazon_v2/}} \cite{NiLM19AmazonReviewDataset}:
`\textit{Industrial and Scientific}' (Scientific), `\textit{Prime Pantry}' (Pantry), `\textit{Musical Instruments}' (Inst.), `\textit{Arts, Crafts \& Sewing}' (Arts), and `\textit{Office Products}' (Office). 
Following \cite{TangWZZZL23MIRACLE, LiWLFSSM23Recformer}, we used 5-core datasets where we filtered out users and items with less than five interactions and items with missing titles.
For zero-shot evaluation, we used a combination of eight categories
\footnote{`\textit{Automotive}', `\textit{Cell Phones and Accessories}', `\textit{Clothing Shoes and Jewelry}', `\textit{Electronics}', `\textit{Grocery and Gourmet Food}', `\textit{Home and Kitchen}', `\textit{Movies and TV}' for training, and `\textit{CDs and Vinyl}' for validation}
for pre-training.
For item attributes, we used title, brand, and category as in \cite{HouMZLDW22UniSRec, LiWLFSSM23Recformer}. 

% \vspace{1mm}
\noindent
\textbf{Baselines and evaluation protocols.} 
We compare MARS with eight baselines: 
\textbf{GRU4Rec}~\cite{HidasiKBT15GRU4Rec}, \textbf{SASRec}~\cite{KangM18SASRec}, and \textbf{BERT4Rec}~\cite{SunLWPLOJ19BERT4Rec} for ID-based models; \ 
\textbf{S$^3$Rec}~\cite{ZhouWZZWZWW20S3Rec} and \textbf{FDSA}~\cite{ZhangZLSXWLZ19FDSA} for ID + Text models; \ 
\textbf{UniSRec}~\cite{HouMZLDW22UniSRec}, \textbf{Recformer}~\cite{LiWLFSSM23Recformer}, and \textbf{MIRACLE}~\cite{TangWZZZL23MIRACLE} for text-based models.
Note that the \textit{inductive setting} is used in UniSRec~\cite{HouMZLDW22UniSRec} and MIRACLE~\cite{TangWZZZL23MIRACLE}, where explicit ID embedding vectors are not used.
Since MARS focuses on the attributes of individual items instead of those that represent relationships among multiple items, we do not consider attributes such as time interval~\cite{LiWM20TISASREC} or co-purchase~\cite{FanLWHPY22MT4SR}.
We adopt Recall@10 (R@10) and NDCG@10 (N@10) for evaluation metrics. We used the \emph{leave-one-out} strategy for train/validation/test data splitting as in~\cite{KangM18SASRec, ZhouWZZWZWW20S3Rec, LiWLFSSM23Recformer}.

% \vspace{1mm}
\noindent
\textbf{Implementation details.}
We initialize the model parameters using Longformer-base~\cite{05150Longformer} where $d=768$ from Huggingface\footnote{\url{https://huggingface.co/allenai/longformer-base-4096}} ~\cite{DBLP:conf/emnlp/WolfDSCDMCRLFDS20Transformers}.
We consider the most recent 50 items for each user. We truncate each attribute value to 32 and the total input sequence to 1,024 tokens.
We set $\tau$ to 0.05. We searched $d'$ from $\{128, 256, 384, 768\}$ and found that $d'=256$ showed a negligible performance degradation while significantly reducing memory footprint.
We used Adam optimizer~\cite{KingmaB14Adam} with a learning rate of 5e-5 and set the batch size as 8 with 16 accumulation steps, 800 warm-up steps, and early stopping based on NDCG@10 on the validation set with patience of 5 epochs.
We implemented the baselines on the open-source library RecBole\footnote{\url{https://github.com/RUCAIBox/RecBole}}~\cite{recbole} or the published source code.
For text-based baselines, we initialize the models with the public pre-trained weight. 

\input{Tables/tab-zero_shot}
% \noindent

\subsection{Results and Analysis}\label{sec:results}
\noindent
\textbf{Overall comparison.} 
Table~\ref{tab:main_table} reports the overall performance comparison between MARS and baselines. 
(i) MARS consistently achieves state-of-the-art performance on all datasets, improving Recall@10 and NDCG@10 by up to 24.43\% and 29.26\%, respectively, against the best competitive baseline.
This indicates that MARS successfully captures the intricate relationships between user sequences and items. 
(ii) MARS constantly outperforms the best competitive text-based model Recformer~\cite{LiWLFSSM23Recformer}, yielding an average gain of 13.79\% and 19.78\% on Recall@10 and NDCG@10.
Since MARS differs from other text-based models in that it considers multiple user/item representations, the performance gain indicates that MARS provides better recommendations by capturing diverse user intents and multi-faceted aspects of items.
(iii) In particular, MARS outperforms MIRACLE~\cite{TangWZZZL23MIRACLE}, which utilizes multiple user representations.
This shows the effectiveness of attribute-wise interaction, while MIRACLE~\cite{TangWZZZL23MIRACLE} computes the score with only one interest through max-pooling.

\input{Tables/tab-ablation}

\noindent
\textbf{Zero-shot recommendation.}\label{sec:zero-shot}
We evaluate the zero-shot recommendation performance of text-based methods to measure knowledge transfer to unseen data.
For a fair evaluation, we pre-train MARS following the setting used in~\cite{LiWLFSSM23Recformer}.
As reported in Table~\ref{tab:zero_shot}, MARS outperforms all baselines with an average gain of 13.11\%, demonstrating the effectiveness of attribute-based encoding and matching in capturing user preferences and transferring knowledge of PLM to a new dataset even without training.

\noindent
\textbf{Ablation study.}\label{sec:ablation}
We compare our proposed attribute-aware representation with two coarse-grained representations: \texttt{[BOS]} (single representation for user/item) and "Item" (average pooling by item).
The results in Table ~\ref{tab:ablation} show that attribute-aware representations yield a performance gain of up to 6.4\% in R@10, highlighting the importance of fine-grained representations.
We also examine the effectiveness of attribute-wise interaction matching, replacing max with mean in Eq.~\eqref{eq:attribute_wise_matching_score}.
The results demonstrate the effectiveness of MaxSim in capturing user preference in an attribute-wise manner.

%% file: Tables/tab-zero_shot.tex
\begin{table}[t] \footnotesize
\centering
\setlength{\tabcolsep}{2pt} 
\setlength{\textfloatsep}{1pt} 
\renewcommand{\arraystretch}{0.8}
\caption{Zero-shot performance for text-based models. The best and second-best are \textbf{bold} and \ul{underlined}. Gain measures the difference between MARS and the best baseline.}
\vspace{-2.5mm}
\label{tab:zero_shot}
\begin{adjustbox}{width=\columnwidth,center}
\begin{tabular}{c|c|cccc|c}
\toprule
{Dataset} & {Metric} & {UniSRec} & {MIRACLE} & {Recformer} & {MARS} & {Gain} \\ \midrule
\multirow{2}{*}{{Scientific}} & {R@10} & 0.0303 & 0.0335 & \ul{0.1296} & \textbf{0.1429} & 10.26\% \\
                                     & {N@10} & 0.0245 & 0.0270 & \ul{0.0850} & \textbf{0.0974} & 14.59\% \\ \midrule
\multirow{2}{*}{{Pantry}}     & {R@10} & 0.0238 & 0.0238 & \ul{0.0663} & \textbf{0.0781} & 17.80\% \\
                                     & {N@10} & 0.0208 & 0.0180 & \ul{0.0379} & \textbf{0.0455} & 20.05\% \\ \midrule
\multirow{2}{*}{{Inst.}}      & {R@10} & 0.0208 & 0.0181 & \ul{0.0714} & \textbf{0.0890} & 24.65\% \\
                                     & {N@10} & 0.0164 & 0.0129 & \ul{0.0414} & \textbf{0.0564} & 36.23\% \\ \midrule
\multirow{2}{*}{{Arts}}       & {R@10} & 0.0330 & 0.0457 & \ul{0.1190} & \textbf{0.1376} & 15.63\% \\
                                     & {N@10} & 0.0264 & 0.0355 & \ul{0.0720} & \textbf{0.0889} & 23.47\% \\ \midrule
\multirow{2}{*}{{Office}}     & {R@10} & 0.0288 & 0.0296 & \ul{0.0847} & \textbf{0.1040} & 22.79\% \\
                                     & {N@10} & 0.0226 & 0.0224 & \ul{0.0547} & \textbf{0.0728} & 33.09\% \\ \bottomrule
\end{tabular}
\end{adjustbox}
\vspace{-5mm}
\end{table}

%% file: Tables/tab-ablation.tex
\begin{table}[t] \footnotesize
\centering
\setlength{\tabcolsep}{2.5pt} 
\renewcommand{\arraystretch}{0.8}
\caption{Ablation study of MARS. The best scores are in bold.}
\vspace{-2mm}
\label{tab:ablation}
\begin{adjustbox}{width=0.8\columnwidth,center}
\begin{tabular}{cc|cc|c|c}
\toprule
\multicolumn{2}{c|}{{Variant}} & \multicolumn{2}{c|}{{Representation}} & {Matching} & {Ours} \\ \midrule
\multicolumn{1}{c|}{{Dataset}} & {Metric} & {\texttt{[BOS]}} & {Item} & {Mean} & {MARS} \\ \midrule
\multicolumn{1}{c|}{\multirow{2}{*}{{Scientific}}} & {R@10} & 0.1373 & { 0.1441} & 0.1420 & \textbf{0.1533} \\
\multicolumn{1}{c|}{} & {N@10} & 0.0991 & { 0.1039} & 0.1016 & \textbf{0.1086} \\ \midrule
\multicolumn{1}{c|}{\multirow{2}{*}{{Pantry}}} & {R@10} & 0.0790 & { 0.0897} & 0.0832 & \textbf{0.0928} \\
\multicolumn{1}{c|}{} & {N@10} & 0.0527 & { 0.0568} & 0.0553 & \textbf{0.0591} \\ \midrule
\multicolumn{1}{c|}{\multirow{2}{*}{{Inst.}}} & {R@10} & 0.0935 & { 0.1046} & 0.0993 & \textbf{0.1082} \\
\multicolumn{1}{c|}{} & {N@10} & 0.0742 & { 0.0823} & 0.0775 & \textbf{0.0846} \\ \midrule
\multicolumn{1}{c|}{\multirow{2}{*}{{Arts}}} & {R@10} & 0.1531 & { 0.1652} & 0.1534 & \textbf{0.1668} \\
\multicolumn{1}{c|}{} & {N@10} & 0.1201 & { 0.1296} & 0.1216 & \textbf{0.1326} \\ \midrule
\multicolumn{1}{c|}{\multirow{2}{*}{{Office}}} & {R@10} & 0.1342 & { 0.1527} & 0.1384 & \textbf{0.1588} \\
\multicolumn{1}{c|}{} & {N@10} & 0.1095 & { 0.1215} & 0.1091 & \textbf{0.1270} \\ \bottomrule
\end{tabular}
\end{adjustbox}
\vspace{-5mm}
\end{table}

%% file: body/4_conclusion.tex
\section{Conclusion}\label{sec:conclusion}

In this paper, we propose MARS, a novel text-based sequential recommender model that fully utilizes fine-grained attribute-aware representations.
MARS employs \textit{multiple attribute-aware representations} by encoding textual attributes of users and items.
Then, the recommendation score is calculated via \textit{attribute-wise item matching}, considering intricate user interests.
Experimental results demonstrate that MARS outperforms competitive baselines in both fine-tuned and zero-shot settings, showing its ability to match fine-grained user sequences and item representations.

%% file: main.bbl
%%% -*-BibTeX-*-
%%% Do NOT edit. File created by BibTeX with style
%%% ACM-Reference-Format-Journals [18-Jan-2012].

\begin{thebibliography}{40}

%%% ====================================================================
%%% NOTE TO THE USER: you can override these defaults by providing
%%% customized versions of any of these macros before the \bibliography
%%% command.  Each of them MUST provide its own final punctuation,
%%% except for \shownote{}, \showDOI{}, and \showURL{}.  The latter two
%%% do not use final punctuation, in order to avoid confusing it with
%%% the Web address.
%%%
%%% To suppress output of a particular field, define its macro to expand
%%% to an empty string, or better, \unskip, like this:
%%%
%%% \newcommand{\showDOI}[1]{\unskip}   % LaTeX syntax
%%%
%%% \def \showDOI #1{\unskip}           % plain TeX syntax
%%%
%%% ====================================================================

\ifx \showCODEN    \undefined \def \showCODEN     #1{\unskip}     \fi
\ifx \showDOI      \undefined \def \showDOI       #1{#1}\fi
\ifx \showISBNx    \undefined \def \showISBNx     #1{\unskip}     \fi
\ifx \showISBNxiii \undefined \def \showISBNxiii  #1{\unskip}     \fi
\ifx \showISSN     \undefined \def \showISSN      #1{\unskip}     \fi
\ifx \showLCCN     \undefined \def \showLCCN      #1{\unskip}     \fi
\ifx \shownote     \undefined \def \shownote      #1{#1}          \fi
\ifx \showarticletitle \undefined \def \showarticletitle #1{#1}   \fi
\ifx \showURL      \undefined \def \showURL       {\relax}        \fi
% The following commands are used for tagged output and should be
% invisible to TeX
\providecommand\bibfield[2]{#2}
\providecommand\bibinfo[2]{#2}
\providecommand\natexlab[1]{#1}
\providecommand\showeprint[2][]{arXiv:#2}

\bibitem[Beltagy et~al\mbox{.}(2020)]%
        {05150Longformer}
\bibfield{author}{\bibinfo{person}{Iz Beltagy}, \bibinfo{person}{Matthew~E. Peters}, {and} \bibinfo{person}{Arman Cohan}.} \bibinfo{year}{2020}\natexlab{}.
\newblock \showarticletitle{Longformer: The Long-Document Transformer}.
\newblock \bibinfo{journal}{\emph{CoRR}} (\bibinfo{year}{2020}).
\newblock


\bibitem[Chang et~al\mbox{.}(2021)]%
        {ChangGZHNSJ021SURGE}
\bibfield{author}{\bibinfo{person}{Jianxin Chang}, \bibinfo{person}{Chen Gao}, \bibinfo{person}{Yu Zheng}, \bibinfo{person}{Yiqun Hui}, \bibinfo{person}{Yanan Niu}, \bibinfo{person}{Yang Song}, \bibinfo{person}{Depeng Jin}, {and} \bibinfo{person}{Yong Li}.} \bibinfo{year}{2021}\natexlab{}.
\newblock \showarticletitle{Sequential Recommendation with Graph Neural Networks}. In \bibinfo{booktitle}{\emph{SIGIR}}. \bibinfo{pages}{378--387}.
\newblock


\bibitem[Chen et~al\mbox{.}(2018a)]%
        {ChenLSZ18Spotify}
\bibfield{author}{\bibinfo{person}{Ching{-}Wei Chen}, \bibinfo{person}{Paul Lamere}, \bibinfo{person}{Markus Schedl}, {and} \bibinfo{person}{Hamed Zamani}.} \bibinfo{year}{2018}\natexlab{a}.
\newblock \showarticletitle{Recsys challenge 2018: automatic music playlist continuation}. In \bibinfo{booktitle}{\emph{{RecSys}}}. \bibinfo{pages}{527--528}.
\newblock


\bibitem[Chen et~al\mbox{.}(2018b)]%
        {ChenXZT0QZ18MANN}
\bibfield{author}{\bibinfo{person}{Xu Chen}, \bibinfo{person}{Hongteng Xu}, \bibinfo{person}{Yongfeng Zhang}, \bibinfo{person}{Jiaxi Tang}, \bibinfo{person}{Yixin Cao}, \bibinfo{person}{Zheng Qin}, {and} \bibinfo{person}{Hongyuan Zha}.} \bibinfo{year}{2018}\natexlab{b}.
\newblock \showarticletitle{Sequential Recommendation with User Memory Networks}. In \bibinfo{booktitle}{\emph{WSDM}}. \bibinfo{pages}{108--116}.
\newblock


\bibitem[Covington et~al\mbox{.}(2016)]%
        {CovingtonAS16}
\bibfield{author}{\bibinfo{person}{Paul Covington}, \bibinfo{person}{Jay Adams}, {and} \bibinfo{person}{Emre Sargin}.} \bibinfo{year}{2016}\natexlab{}.
\newblock \showarticletitle{Deep Neural Networks for YouTube Recommendations}. In \bibinfo{booktitle}{\emph{RecSys}}. \bibinfo{pages}{191--198}.
\newblock


\bibitem[Ding et~al\mbox{.}(2021)]%
        {Hao21ZesRec}
\bibfield{author}{\bibinfo{person}{Hao Ding}, \bibinfo{person}{Yifei Ma}, \bibinfo{person}{Anoop Deoras}, \bibinfo{person}{Yuyang Wang}, {and} \bibinfo{person}{Hao Wang}.} \bibinfo{year}{2021}\natexlab{}.
\newblock \showarticletitle{Zero-Shot Recommender Systems}.
\newblock \bibinfo{journal}{\emph{CoRR}} (\bibinfo{year}{2021}).
\newblock


\bibitem[Du et~al\mbox{.}(2022)]%
        {DuSZWSLL022CBIT}
\bibfield{author}{\bibinfo{person}{Hanwen Du}, \bibinfo{person}{Hui Shi}, \bibinfo{person}{Pengpeng Zhao}, \bibinfo{person}{Deqing Wang}, \bibinfo{person}{Victor~S. Sheng}, \bibinfo{person}{Yanchi Liu}, \bibinfo{person}{Guanfeng Liu}, {and} \bibinfo{person}{Lei Zhao}.} \bibinfo{year}{2022}\natexlab{}.
\newblock \showarticletitle{Contrastive Learning with Bidirectional Transformers for Sequential Recommendation}. In \bibinfo{booktitle}{\emph{CIKM}}. \bibinfo{pages}{396--405}.
\newblock


\bibitem[Fan et~al\mbox{.}({[n.\,d.]})]%
        {FanLWHPY22MT4SR}
\bibfield{author}{\bibinfo{person}{Ziwei Fan}, \bibinfo{person}{Zhiwei Liu}, \bibinfo{person}{Chen Wang}, \bibinfo{person}{Peijie Huang}, \bibinfo{person}{Hao Peng}, {and} \bibinfo{person}{Philip~S. Yu}.} \bibinfo{year}{[n.\,d.]}\natexlab{}.
\newblock \showarticletitle{Sequential Recommendation with Auxiliary Item Relationships via Multi-Relational Transformer}. In \bibinfo{booktitle}{\emph{Big Data}}. \bibinfo{pages}{525--534}.
\newblock


\bibitem[Fang et~al\mbox{.}(2019)]%
        {FangGZS19Survey}
\bibfield{author}{\bibinfo{person}{Hui Fang}, \bibinfo{person}{Guibing Guo}, \bibinfo{person}{Danning Zhang}, {and} \bibinfo{person}{Yiheng Shu}.} \bibinfo{year}{2019}\natexlab{}.
\newblock \showarticletitle{Deep Learning-Based Sequential Recommender Systems: Concepts, Algorithms, and Evaluations}. In \bibinfo{booktitle}{\emph{ICWE}}, Vol.~\bibinfo{volume}{11496}. \bibinfo{pages}{574--577}.
\newblock


\bibitem[Fang et~al\mbox{.}(2020)]%
        {FangZSG20SeqSurvey}
\bibfield{author}{\bibinfo{person}{Hui Fang}, \bibinfo{person}{Danning Zhang}, \bibinfo{person}{Yiheng Shu}, {and} \bibinfo{person}{Guibing Guo}.} \bibinfo{year}{2020}\natexlab{}.
\newblock \showarticletitle{Deep Learning for Sequential Recommendation: Algorithms, Influential Factors, and Evaluations}.
\newblock \bibinfo{journal}{\emph{{ACM} Trans. Inf. Syst.}} \bibinfo{volume}{39}, \bibinfo{number}{1} (\bibinfo{year}{2020}), \bibinfo{pages}{10:1--10:42}.
\newblock


\bibitem[Gupta et~al\mbox{.}(2019)]%
        {GuptaGMVS19NISER}
\bibfield{author}{\bibinfo{person}{Priyanka Gupta}, \bibinfo{person}{Diksha Garg}, \bibinfo{person}{Pankaj Malhotra}, \bibinfo{person}{Lovekesh Vig}, {and} \bibinfo{person}{Gautam~M. Shroff}.} \bibinfo{year}{2019}\natexlab{}.
\newblock \showarticletitle{{NISER:} Normalized Item and Session Representations with Graph Neural Networks}.
\newblock \bibinfo{journal}{\emph{CoRR}} (\bibinfo{year}{2019}).
\newblock


\bibitem[Hidasi et~al\mbox{.}(2016)]%
        {HidasiKBT15GRU4Rec}
\bibfield{author}{\bibinfo{person}{Bal{\'{a}}zs Hidasi}, \bibinfo{person}{Alexandros Karatzoglou}, \bibinfo{person}{Linas Baltrunas}, {and} \bibinfo{person}{Domonkos Tikk}.} \bibinfo{year}{2016}\natexlab{}.
\newblock \showarticletitle{Session-based Recommendations with Recurrent Neural Networks}. In \bibinfo{booktitle}{\emph{ICLR}}.
\newblock


\bibitem[Hinton et~al\mbox{.}(2015)]%
        {HintonVD15distillation}
\bibfield{author}{\bibinfo{person}{Geoffrey~E. Hinton}, \bibinfo{person}{Oriol Vinyals}, {and} \bibinfo{person}{Jeffrey Dean}.} \bibinfo{year}{2015}\natexlab{}.
\newblock \showarticletitle{Distilling the Knowledge in a Neural Network}.
\newblock \bibinfo{journal}{\emph{CoRR}} (\bibinfo{year}{2015}).
\newblock


\bibitem[Hou et~al\mbox{.}(2022)]%
        {HouMZLDW22UniSRec}
\bibfield{author}{\bibinfo{person}{Yupeng Hou}, \bibinfo{person}{Shanlei Mu}, \bibinfo{person}{Wayne~Xin Zhao}, \bibinfo{person}{Yaliang Li}, \bibinfo{person}{Bolin Ding}, {and} \bibinfo{person}{Ji{-}Rong Wen}.} \bibinfo{year}{2022}\natexlab{}.
\newblock \showarticletitle{Towards Universal Sequence Representation Learning for Recommender Systems}. In \bibinfo{booktitle}{\emph{KDD}}. \bibinfo{pages}{585--593}.
\newblock


\bibitem[Kang and McAuley(2018)]%
        {KangM18SASRec}
\bibfield{author}{\bibinfo{person}{Wang{-}Cheng Kang} {and} \bibinfo{person}{Julian~J. McAuley}.} \bibinfo{year}{2018}\natexlab{}.
\newblock \showarticletitle{Self-Attentive Sequential Recommendation}. In \bibinfo{booktitle}{\emph{ICDM}}. \bibinfo{pages}{197--206}.
\newblock


\bibitem[Khattab and Zaharia(2020)]%
        {KhattabZ20ColBERT}
\bibfield{author}{\bibinfo{person}{Omar Khattab} {and} \bibinfo{person}{Matei Zaharia}.} \bibinfo{year}{2020}\natexlab{}.
\newblock \showarticletitle{ColBERT: Efficient and Effective Passage Search via Contextualized Late Interaction over {BERT}}. In \bibinfo{booktitle}{\emph{SIGIR}}. \bibinfo{pages}{39--48}.
\newblock


\bibitem[Kingma and Ba(2015)]%
        {KingmaB14Adam}
\bibfield{author}{\bibinfo{person}{Diederik~P. Kingma} {and} \bibinfo{person}{Jimmy Ba}.} \bibinfo{year}{2015}\natexlab{}.
\newblock \showarticletitle{Adam: {A} Method for Stochastic Optimization}. In \bibinfo{booktitle}{\emph{ICLR}}.
\newblock


\bibitem[Li et~al\mbox{.}(2022)]%
        {LiZZYCSKN22RecGURU}
\bibfield{author}{\bibinfo{person}{Chenglin Li}, \bibinfo{person}{Mingjun Zhao}, \bibinfo{person}{Huanming Zhang}, \bibinfo{person}{Chenyun Yu}, \bibinfo{person}{Lei Cheng}, \bibinfo{person}{Guoqiang Shu}, \bibinfo{person}{Beibei Kong}, {and} \bibinfo{person}{Di Niu}.} \bibinfo{year}{2022}\natexlab{}.
\newblock \showarticletitle{RecGURU: Adversarial Learning of Generalized User Representations for Cross-Domain Recommendation}. In \bibinfo{booktitle}{\emph{WSDM}}. \bibinfo{pages}{571--581}.
\newblock


\bibitem[Li et~al\mbox{.}(2017)]%
        {LiRCRLM17NARM}
\bibfield{author}{\bibinfo{person}{Jing Li}, \bibinfo{person}{Pengjie Ren}, \bibinfo{person}{Zhumin Chen}, \bibinfo{person}{Zhaochun Ren}, \bibinfo{person}{Tao Lian}, {and} \bibinfo{person}{Jun Ma}.} \bibinfo{year}{2017}\natexlab{}.
\newblock \showarticletitle{Neural Attentive Session-based Recommendation}. In \bibinfo{booktitle}{\emph{CIKM}}. \bibinfo{pages}{1419--1428}.
\newblock


\bibitem[Li et~al\mbox{.}(2023)]%
        {LiWLFSSM23Recformer}
\bibfield{author}{\bibinfo{person}{Jiacheng Li}, \bibinfo{person}{Ming Wang}, \bibinfo{person}{Jin Li}, \bibinfo{person}{Jinmiao Fu}, \bibinfo{person}{Xin Shen}, \bibinfo{person}{Jingbo Shang}, {and} \bibinfo{person}{Julian~J. McAuley}.} \bibinfo{year}{2023}\natexlab{}.
\newblock \showarticletitle{Text Is All You Need: Learning Language Representations for Sequential Recommendation}. In \bibinfo{booktitle}{\emph{KDD}}. \bibinfo{pages}{1258--1267}.
\newblock


\bibitem[Li et~al\mbox{.}(2020)]%
        {LiWM20TISASREC}
\bibfield{author}{\bibinfo{person}{Jiacheng Li}, \bibinfo{person}{Yujie Wang}, {and} \bibinfo{person}{Julian~J. McAuley}.} \bibinfo{year}{2020}\natexlab{}.
\newblock \showarticletitle{Time Interval Aware Self-Attention for Sequential Recommendation}. In \bibinfo{booktitle}{\emph{WSDM}}. \bibinfo{pages}{322--330}.
\newblock


\bibitem[Linden et~al\mbox{.}(2003)]%
        {LindenSY03Amazon}
\bibfield{author}{\bibinfo{person}{Greg Linden}, \bibinfo{person}{Brent Smith}, {and} \bibinfo{person}{Jeremy York}.} \bibinfo{year}{2003}\natexlab{}.
\newblock \showarticletitle{Amazon.com Recommendations: Item-to-Item Collaborative Filtering}.
\newblock \bibinfo{journal}{\emph{{IEEE} Internet Comput.}} \bibinfo{volume}{7}, \bibinfo{number}{1} (\bibinfo{year}{2003}), \bibinfo{pages}{76--80}.
\newblock


\bibitem[Liu et~al\mbox{.}(2021)]%
        {Liu21NOVA}
\bibfield{author}{\bibinfo{person}{Chang Liu}, \bibinfo{person}{Xiaoguang Li}, \bibinfo{person}{Guohao Cai}, \bibinfo{person}{Zhenhua Dong}, \bibinfo{person}{Hong Zhu}, {and} \bibinfo{person}{Lifeng Shang}.} \bibinfo{year}{2021}\natexlab{}.
\newblock \showarticletitle{Non-invasive Self-attention for Side Information Fusion in Sequential Recommendation}. In \bibinfo{booktitle}{\emph{AAAI}}. \bibinfo{pages}{4249--4256}.
\newblock


\bibitem[Liu et~al\mbox{.}(2023)]%
        {LiuMXLYL0023TASTE}
\bibfield{author}{\bibinfo{person}{Zhenghao Liu}, \bibinfo{person}{Sen Mei}, \bibinfo{person}{Chenyan Xiong}, \bibinfo{person}{Xiaohua Li}, \bibinfo{person}{Shi Yu}, \bibinfo{person}{Zhiyuan Liu}, \bibinfo{person}{Yu Gu}, {and} \bibinfo{person}{Ge Yu}.} \bibinfo{year}{2023}\natexlab{}.
\newblock \showarticletitle{Text Matching Improves Sequential Recommendation by Reducing Popularity Biases}. In \bibinfo{booktitle}{\emph{CIKM}}. \bibinfo{pages}{1534--1544}.
\newblock


\bibitem[Ni et~al\mbox{.}(2019)]%
        {NiLM19AmazonReviewDataset}
\bibfield{author}{\bibinfo{person}{Jianmo Ni}, \bibinfo{person}{Jiacheng Li}, {and} \bibinfo{person}{Julian~J. McAuley}.} \bibinfo{year}{2019}\natexlab{}.
\newblock \showarticletitle{Justifying Recommendations using Distantly-Labeled Reviews and Fine-Grained Aspects}. In \bibinfo{booktitle}{\emph{EMNLP-IJCNLP}}. \bibinfo{pages}{188--197}.
\newblock


\bibitem[Qiu et~al\mbox{.}(2022)]%
        {QiuHYW22DuoRec}
\bibfield{author}{\bibinfo{person}{Ruihong Qiu}, \bibinfo{person}{Zi Huang}, \bibinfo{person}{Hongzhi Yin}, {and} \bibinfo{person}{Zijian Wang}.} \bibinfo{year}{2022}\natexlab{}.
\newblock \showarticletitle{Contrastive Learning for Representation Degeneration Problem in Sequential Recommendation}. In \bibinfo{booktitle}{\emph{WSDM}}. \bibinfo{pages}{813--823}.
\newblock


\bibitem[Sun et~al\mbox{.}(2019)]%
        {SunLWPLOJ19BERT4Rec}
\bibfield{author}{\bibinfo{person}{Fei Sun}, \bibinfo{person}{Jun Liu}, \bibinfo{person}{Jian Wu}, \bibinfo{person}{Changhua Pei}, \bibinfo{person}{Xiao Lin}, \bibinfo{person}{Wenwu Ou}, {and} \bibinfo{person}{Peng Jiang}.} \bibinfo{year}{2019}\natexlab{}.
\newblock \showarticletitle{BERT4Rec: Sequential Recommendation with Bidirectional Encoder Representations from Transformer}. In \bibinfo{booktitle}{\emph{CIKM}}. \bibinfo{pages}{1441--1450}.
\newblock


\bibitem[Tang and Wang(2018)]%
        {TangW18caser}
\bibfield{author}{\bibinfo{person}{Jiaxi Tang} {and} \bibinfo{person}{Ke Wang}.} \bibinfo{year}{2018}\natexlab{}.
\newblock \showarticletitle{Personalized Top-N Sequential Recommendation via Convolutional Sequence Embedding}. In \bibinfo{booktitle}{\emph{WSDM}}. \bibinfo{publisher}{{ACM}}, \bibinfo{pages}{565--573}.
\newblock


\bibitem[Tang et~al\mbox{.}(2023)]%
        {TangWZZZL23MIRACLE}
\bibfield{author}{\bibinfo{person}{Zuoli Tang}, \bibinfo{person}{Lin Wang}, \bibinfo{person}{Lixin Zou}, \bibinfo{person}{Xiaolu Zhang}, \bibinfo{person}{Jun Zhou}, {and} \bibinfo{person}{Chenliang Li}.} \bibinfo{year}{2023}\natexlab{}.
\newblock \showarticletitle{Towards Multi-Interest Pre-training with Sparse Capsule Network}. In \bibinfo{booktitle}{\emph{SIGIR}}. \bibinfo{pages}{311--320}.
\newblock


\bibitem[Wang et~al\mbox{.}(2019)]%
        {WangHWCSO19SeqSurvey}
\bibfield{author}{\bibinfo{person}{Shoujin Wang}, \bibinfo{person}{Liang Hu}, \bibinfo{person}{Yan Wang}, \bibinfo{person}{Longbing Cao}, \bibinfo{person}{Quan~Z. Sheng}, {and} \bibinfo{person}{Mehmet~A. Orgun}.} \bibinfo{year}{2019}\natexlab{}.
\newblock \showarticletitle{Sequential Recommender Systems: Challenges, Progress and Prospects}. In \bibinfo{booktitle}{\emph{IJCAI}}. \bibinfo{pages}{6332--6338}.
\newblock


\bibitem[Wolf et~al\mbox{.}(2020)]%
        {DBLP:conf/emnlp/WolfDSCDMCRLFDS20Transformers}
\bibfield{author}{\bibinfo{person}{Thomas Wolf}, \bibinfo{person}{Lysandre Debut}, \bibinfo{person}{Victor Sanh}, \bibinfo{person}{Julien Chaumond}, \bibinfo{person}{Clement Delangue}, \bibinfo{person}{Anthony Moi}, \bibinfo{person}{Pierric Cistac}, \bibinfo{person}{Tim Rault}, \bibinfo{person}{R{\'{e}}mi Louf}, \bibinfo{person}{Morgan Funtowicz}, \bibinfo{person}{Joe Davison}, \bibinfo{person}{Sam Shleifer}, \bibinfo{person}{Patrick von Platen}, \bibinfo{person}{Clara Ma}, \bibinfo{person}{Yacine Jernite}, \bibinfo{person}{Julien Plu}, \bibinfo{person}{Canwen Xu}, \bibinfo{person}{Teven~Le Scao}, \bibinfo{person}{Sylvain Gugger}, \bibinfo{person}{Mariama Drame}, \bibinfo{person}{Quentin Lhoest}, {and} \bibinfo{person}{Alexander~M. Rush}.} \bibinfo{year}{2020}\natexlab{}.
\newblock \showarticletitle{Transformers: State-of-the-Art Natural Language Processing}. In \bibinfo{booktitle}{\emph{{EMNLP} (Demos)}}. \bibinfo{publisher}{Association for Computational Linguistics}, \bibinfo{pages}{38--45}.
\newblock


\bibitem[Wu et~al\mbox{.}(2020)]%
        {WuLHS20SSEPT}
\bibfield{author}{\bibinfo{person}{Liwei Wu}, \bibinfo{person}{Shuqing Li}, \bibinfo{person}{Cho{-}Jui Hsieh}, {and} \bibinfo{person}{James Sharpnack}.} \bibinfo{year}{2020}\natexlab{}.
\newblock \showarticletitle{{SSE-PT:} Sequential Recommendation Via Personalized Transformer}. In \bibinfo{booktitle}{\emph{RecSys}}. \bibinfo{pages}{328--337}.
\newblock


\bibitem[Wu et~al\mbox{.}(2019)]%
        {WuT0WXT19SRGNN}
\bibfield{author}{\bibinfo{person}{Shu Wu}, \bibinfo{person}{Yuyuan Tang}, \bibinfo{person}{Yanqiao Zhu}, \bibinfo{person}{Liang Wang}, \bibinfo{person}{Xing Xie}, {and} \bibinfo{person}{Tieniu Tan}.} \bibinfo{year}{2019}\natexlab{}.
\newblock \showarticletitle{Session-Based Recommendation with Graph Neural Networks}. In \bibinfo{booktitle}{\emph{AAAI}}. \bibinfo{pages}{346--353}.
\newblock


\bibitem[Xie et~al\mbox{.}(2022a)]%
        {XieSLWGZDC22CL4SRec}
\bibfield{author}{\bibinfo{person}{Xu Xie}, \bibinfo{person}{Fei Sun}, \bibinfo{person}{Zhaoyang Liu}, \bibinfo{person}{Shiwen Wu}, \bibinfo{person}{Jinyang Gao}, \bibinfo{person}{Jiandong Zhang}, \bibinfo{person}{Bolin Ding}, {and} \bibinfo{person}{Bin Cui}.} \bibinfo{year}{2022}\natexlab{a}.
\newblock \showarticletitle{Contrastive Learning for Sequential Recommendation}. In \bibinfo{booktitle}{\emph{ICDE}}. \bibinfo{pages}{1259--1273}.
\newblock


\bibitem[Xie et~al\mbox{.}(2022b)]%
        {XieZK22DIFSR}
\bibfield{author}{\bibinfo{person}{Yueqi Xie}, \bibinfo{person}{Peilin Zhou}, {and} \bibinfo{person}{Sunghun Kim}.} \bibinfo{year}{2022}\natexlab{b}.
\newblock \showarticletitle{Decoupled Side Information Fusion for Sequential Recommendation}. In \bibinfo{booktitle}{\emph{SIGIR}}.
\newblock


\bibitem[Yuan et~al\mbox{.}(2021)]%
        {YuanDTSZ21ICAISR}
\bibfield{author}{\bibinfo{person}{Xu Yuan}, \bibinfo{person}{Dongsheng Duan}, \bibinfo{person}{Lingling Tong}, \bibinfo{person}{Lei Shi}, {and} \bibinfo{person}{Cheng Zhang}.} \bibinfo{year}{2021}\natexlab{}.
\newblock \showarticletitle{{ICAI-SR:} Item Categorical Attribute Integrated Sequential Recommendation}. In \bibinfo{booktitle}{\emph{SIGIR}}. \bibinfo{pages}{1687--1691}.
\newblock


\bibitem[Zhang et~al\mbox{.}(2019)]%
        {ZhangZLSXWLZ19FDSA}
\bibfield{author}{\bibinfo{person}{Tingting Zhang}, \bibinfo{person}{Pengpeng Zhao}, \bibinfo{person}{Yanchi Liu}, \bibinfo{person}{Victor~S. Sheng}, \bibinfo{person}{Jiajie Xu}, \bibinfo{person}{Deqing Wang}, \bibinfo{person}{Guanfeng Liu}, {and} \bibinfo{person}{Xiaofang Zhou}.} \bibinfo{year}{2019}\natexlab{}.
\newblock \showarticletitle{Feature-level Deeper Self-Attention Network for Sequential Recommendation}. In \bibinfo{booktitle}{\emph{IJCAI}}. \bibinfo{pages}{4320--4326}.
\newblock


\bibitem[Zhao et~al\mbox{.}(2021)]%
        {recbole}
\bibfield{author}{\bibinfo{person}{Wayne~Xin Zhao}, \bibinfo{person}{Shanlei Mu}, \bibinfo{person}{Yupeng Hou}, \bibinfo{person}{Zihan Lin}, \bibinfo{person}{Yushuo Chen}, \bibinfo{person}{Xingyu Pan}, \bibinfo{person}{Kaiyuan Li}, \bibinfo{person}{Yujie Lu}, \bibinfo{person}{Hui Wang}, \bibinfo{person}{Changxin Tian}, \bibinfo{person}{Yingqian Min}, \bibinfo{person}{Zhichao Feng}, \bibinfo{person}{Xinyan Fan}, \bibinfo{person}{Xu Chen}, \bibinfo{person}{Pengfei Wang}, \bibinfo{person}{Wendi Ji}, \bibinfo{person}{Yaliang Li}, \bibinfo{person}{Xiaoling Wang}, {and} \bibinfo{person}{Ji{-}Rong Wen}.} \bibinfo{year}{2021}\natexlab{}.
\newblock \showarticletitle{RecBole: Towards a Unified, Comprehensive and Efficient Framework for Recommendation Algorithms}. In \bibinfo{booktitle}{\emph{{CIKM}}}. \bibinfo{publisher}{{ACM}}, \bibinfo{pages}{4653--4664}.
\newblock


\bibitem[Zhou et~al\mbox{.}(2020)]%
        {ZhouWZZWZWW20S3Rec}
\bibfield{author}{\bibinfo{person}{Kun Zhou}, \bibinfo{person}{Hui Wang}, \bibinfo{person}{Wayne~Xin Zhao}, \bibinfo{person}{Yutao Zhu}, \bibinfo{person}{Sirui Wang}, \bibinfo{person}{Fuzheng Zhang}, \bibinfo{person}{Zhongyuan Wang}, {and} \bibinfo{person}{Ji{-}Rong Wen}.} \bibinfo{year}{2020}\natexlab{}.
\newblock \showarticletitle{S3-Rec: Self-Supervised Learning for Sequential Recommendation with Mutual Information Maximization}. In \bibinfo{booktitle}{\emph{CIKM}}. \bibinfo{pages}{1893--1902}.
\newblock


\bibitem[Zhou et~al\mbox{.}(2022)]%
        {ZhouYZW22FMLPRec}
\bibfield{author}{\bibinfo{person}{Kun Zhou}, \bibinfo{person}{Hui Yu}, \bibinfo{person}{Wayne~Xin Zhao}, {and} \bibinfo{person}{Ji{-}Rong Wen}.} \bibinfo{year}{2022}\natexlab{}.
\newblock \showarticletitle{Filter-enhanced {MLP} is All You Need for Sequential Recommendation}. In \bibinfo{booktitle}{\emph{WWW}}. \bibinfo{pages}{2388--2399}.
\newblock


\end{thebibliography}
